\documentclass{desyproc}
\usepackage{epstopdf}
\usepackage{subfigure}
\usepackage{lineno}

\begin{document}
%------------------------------------
\title{Results of Soft-Diffraction at LHCb}

%for single authors the superscripts are optional
\author{{\slshape Marco Meissner$^1$, on behalf of the LHCb collaboration}\\[1ex]
$^1$Physikalisches Institut, Ruprecht-Karls-Universit\"at Heidelberg, INF 226, 69120 Heidelberg, Germany\\ }

% if the proceedings are available online (e.g. at Indico)
% please enter the contribution ID or file_name below for the DOI
%\contribID{32}
\contribID{meissner\_marco}

% TO THE CONFERENCE EDITORS: 
% please update the following information      
% before sending the template to the authors
% \confID{800}  % if the conference is on Indico uncomment this line

\acronym{EDS'13} % if you want the Acronym in the page footer uncomment this line

\maketitle

\begin{abstract}
The LHCb detector with its unique pseudorapidity coverage allows to perform soft-QCD measurements in the kinematic forward region where QCD models have large uncertainties. Selected analyses related to soft-Diffraction will be summarised in these proceedings. Energy flow and charged particle multiplicity have been measured separately in different event classes. They give input for modelling the underlying event in \textit{pp} collisions. Prompt hadron ratios are important for hadronisation models, while the $\overline p/p$ ratio is a good observable to test models of baryon number transport.
\end{abstract}

\section{Introduction}
The LHCb experiment at the Large Hadron Collider is a dedicated experiment to study CP-violating processes and rare decays of hadrons containing beauty and charm quarks. The detector is a single-arm forward spectrometer~\cite{LHCb} designed to efficiently detect the decay products of B-hadrons in a pseudorapidity range of approximately $2<\eta<5$. This also allows LHCb to make soft-QCD measurements in a kinematic region which is hardly accessible by the general purpose detectors. The analyses presented in these proceedings are selected soft-QCD measurements in the context of soft-diffraction. The data used for these analyses are Proton-Proton (\textit{pp}) collisions at centre-of-mass energies of $\sqrt{s}=0.9$ and $7\;$TeV recorded with minimum bias triggers in the low luminosity running phase in 2010.
Important for the presented analyses are the tracking system, which is composed of a high precision Silicon Vertex Locator (VELO) surrounding the interaction point and the main tracking stations located downstream of a dipole magnet. Particle identification is performed by two Ring Imaging Cherenkov (RICH) detectors which allow separation of charged particles in a momentum range of $2-100\;$GeV/c.

\section{Forward Energy Flow}
For a particular pseudorapidity interval $\Delta\eta$ the total energy flow is defined as 
\begin{equation}
 \frac{1}{N_{int}} \frac{dE_{tot}}{d\eta}=\frac{1}{\Delta\eta}\left( \frac{1}{N_{int}}\sum_{i=1}^{N_{part,\eta}} E_{i,\eta} \right),
\end{equation}
where $N_{int}$ is the number of inelastic \textit{pp} interactions. Measuring the energy flow (EF) at large pseudorapidities directly probes multi-parton interactions (MPI) and parton radiation which contribute to the underlying event in proton-proton collisions. The measurement has been performed in 4 different event classes, an (1) inclusive minimum bias sample which requires to have at least one reconstructed track with a momentum $p$ greater than $2\;$GeV/c in the forward acceptance ($1.9<\eta<4.9$). Further there is a (2) hard scattering sub-sample which implies at least one high $p_{T}$ track per event ($p_{T}>3\;$GeV/c). By exploiting the additional backwards coverage of the VELO it was possible to obtain a (3) diffractive enriched and a (4) non-diffractive enriched sample of events. These were selected by looking for backward tracks in the pseudorapidity range of $-3.5<\eta<-1.5$. This selection exploits the fact, that a large rapidity gap is an experimental signature to identify diffractive processes.
\begin{wrapfigure}{l}{0.61\textwidth}
\vspace{-5pt}
\includegraphics[width=0.3\textwidth]{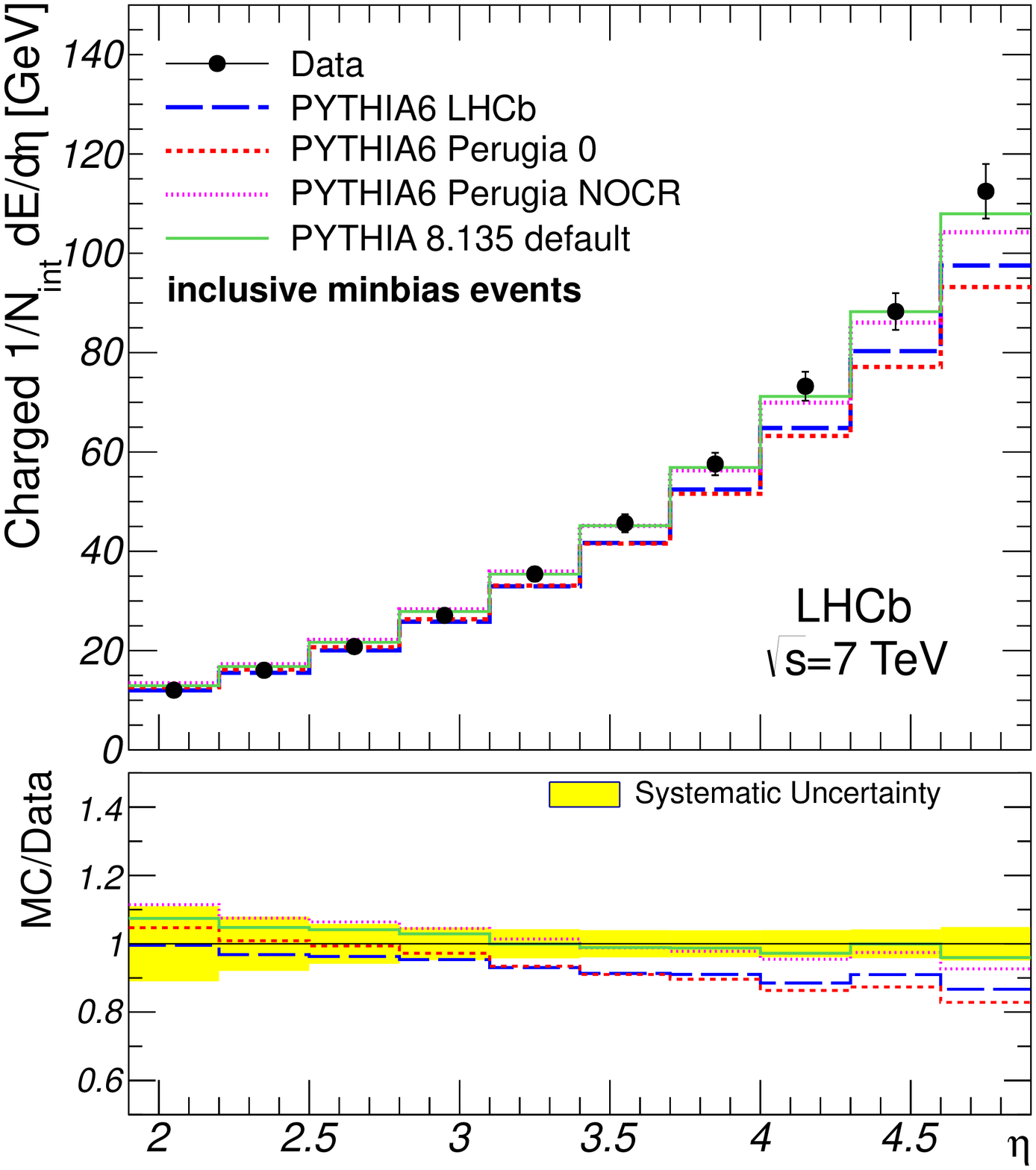}
\includegraphics[width=0.3\textwidth]{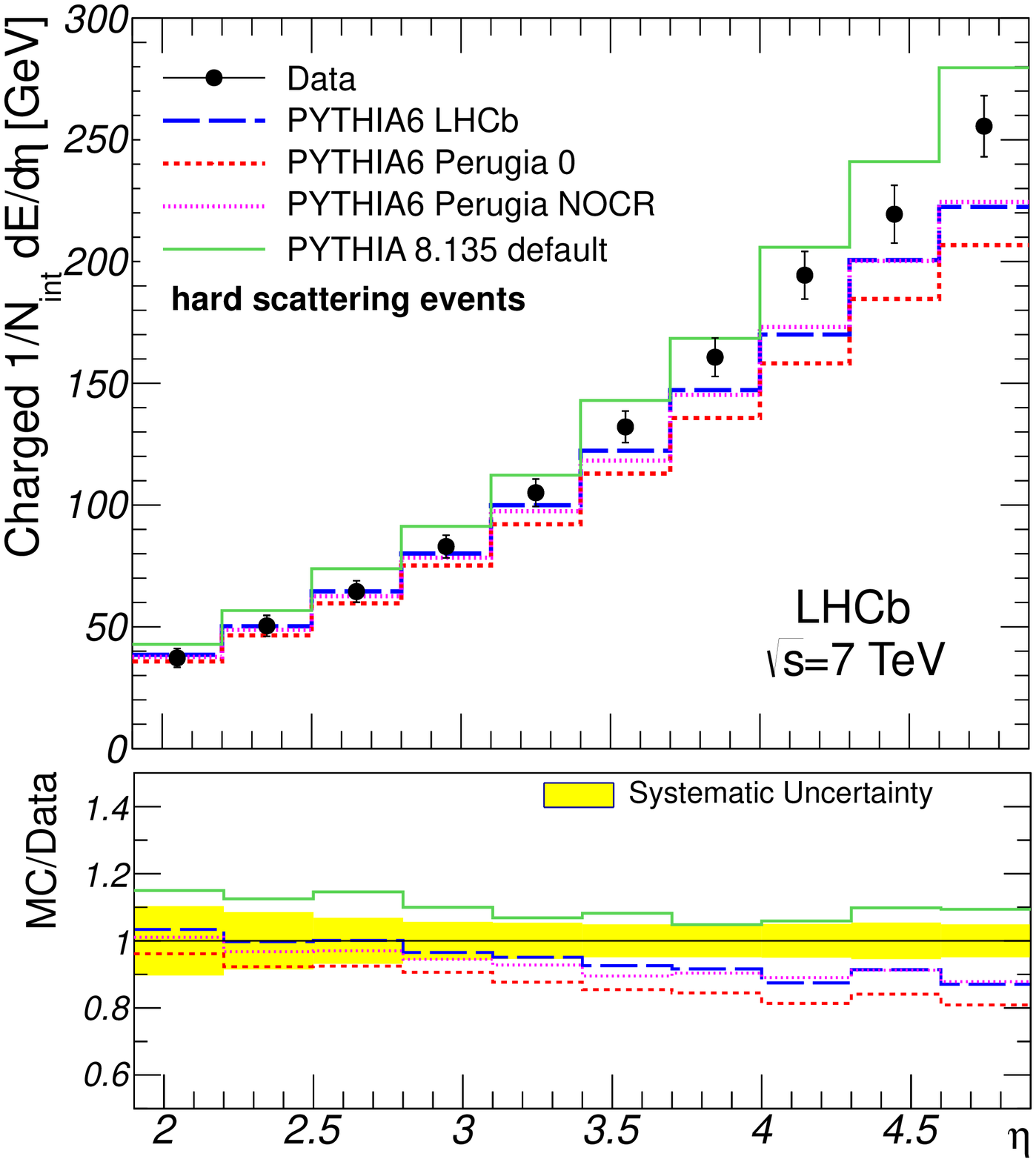}\\
\includegraphics[width=0.3\textwidth]{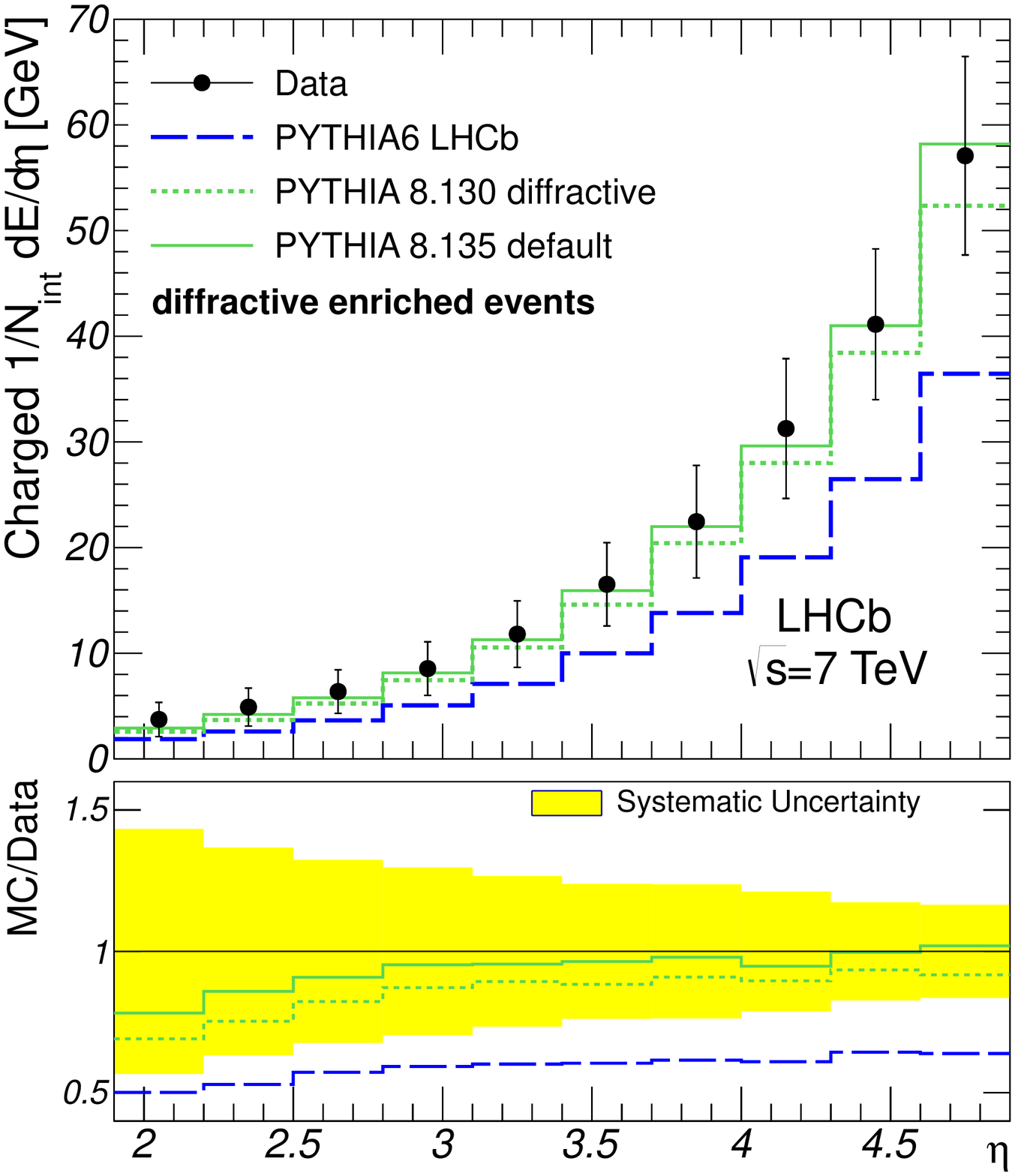}
\includegraphics[width=0.3\textwidth]{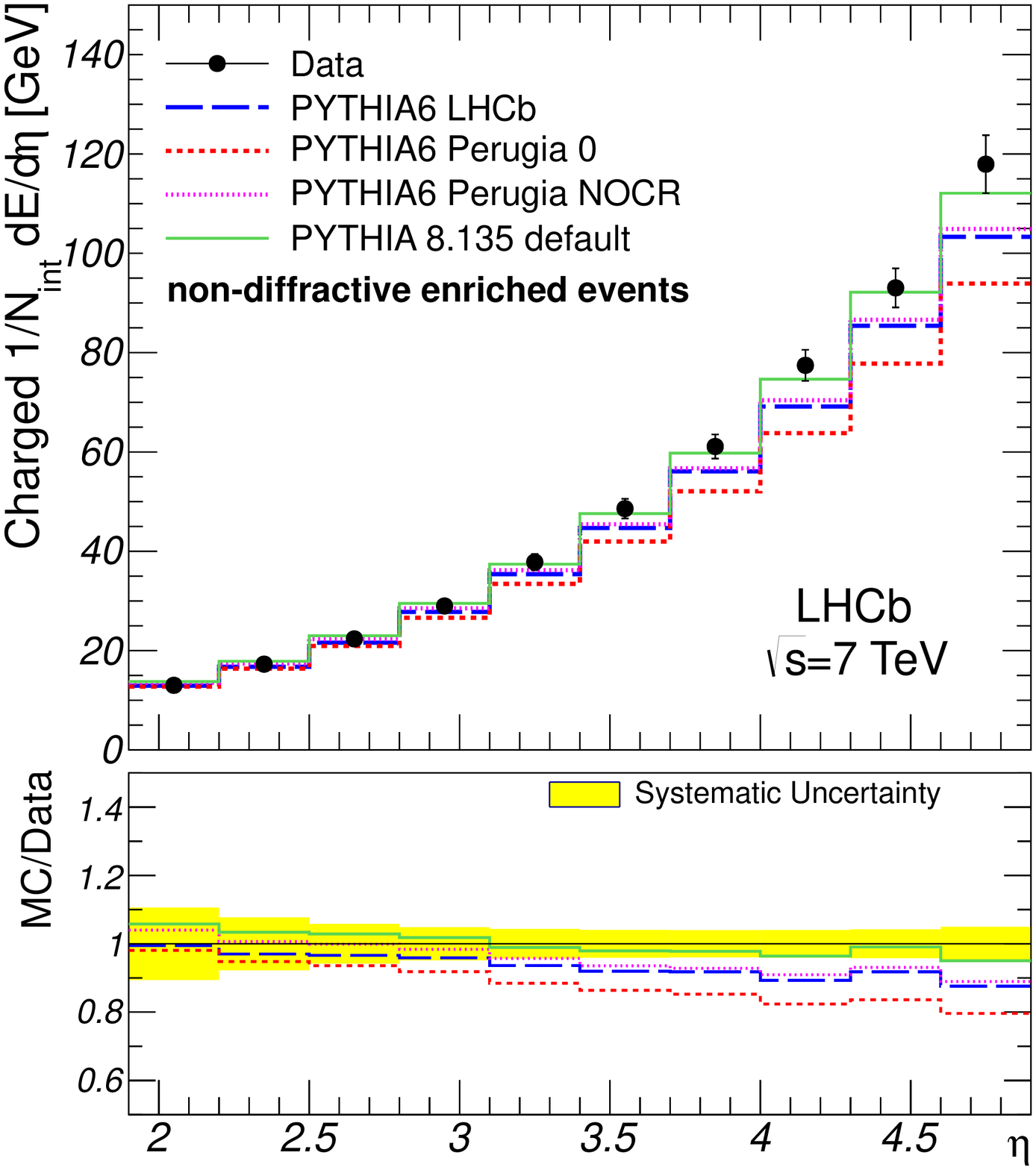}
\caption{Total energy flow as function of $\eta$. LHCb data compared to PYTHIA predictions in four event classes.}\label{EF_1}
\vspace{-5pt}
\end{wrapfigure}
The measured total EF, which is the sum of charged and neutral EF, is depicted in Fig.\ref{EF_1}, superimposed with different PYTHIA generator predictions. The EF in the four event samples increases from the diffractive sample to the inclusive minimum bias and non-diffractive sample up to the hard scattering sample. The errors are dominated by systematic uncertainties, like model dependence for correcting detector effects, uncertainties for the track finding and residual pile-up. These uncertainties decrease towards larger $\eta$ which is the most interesting region for studying MPI phenomena.
In all event classes, the PYTHIA 6 tunes underestimate the EF especially at larger pseudorapidities but overestimate at lower $\eta$. The default PYTHIA 8 prediction (8.135) is in better agreement except for the hard scattering sample. The energy flow in diffractive enriched events is well described by PYTHIA 8. The measurement was also compared to predictions of cosmic ray generators (details see~\cite{LHCb_EF}) which were not tuned to LHC data. The EPOS and SYBILL generators show a good agreement with data in the minimum bias and non-diffractive sample while QGSJET predictions are best for hard scattering. The EF in diffractive events seem to be underestimated by all cosmic ray generators.

\section{Charged Particle Multiplicity}
The multiplicity of primary produced charged particles has been measured~\cite{LHCb_CPM} for \textit{pp} collisions at $\sqrt{s}=7\;$TeV. Primary particles are defined as either directly produced in the \textit{pp} collision or from short lived decays ($\tau<10\;ps$). For this measurement only information from the VELO has been used. 
As there is a negligible influence from the magnetic field in this sub-detector the measurement has no explicit momentum cut-off for low energetic particles. 
\begin{wrapfigure}{r}{0.61\textwidth} 
\vspace{-10pt}
\includegraphics[width=0.3\textwidth]{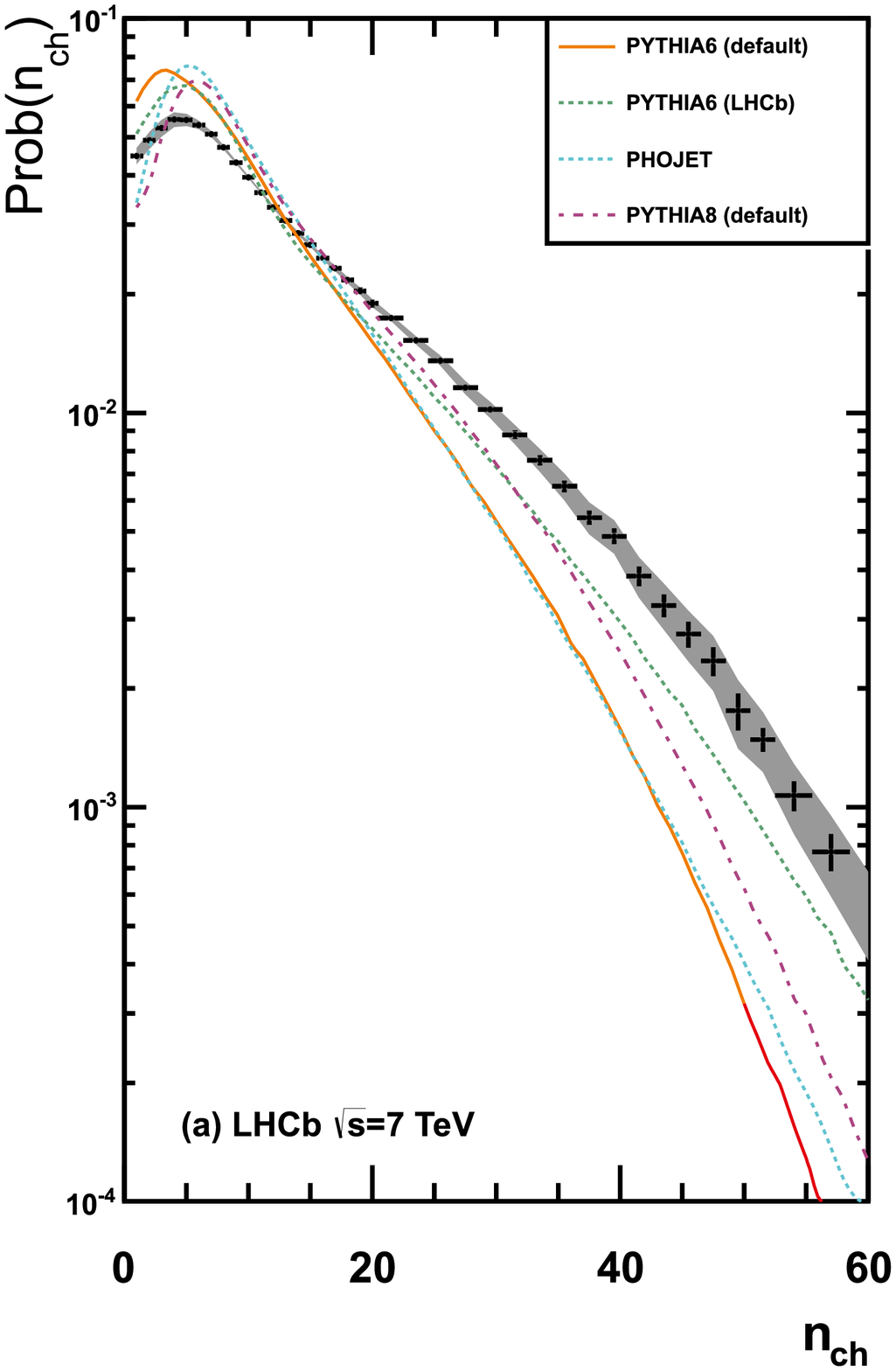}
\includegraphics[width=0.3\textwidth]{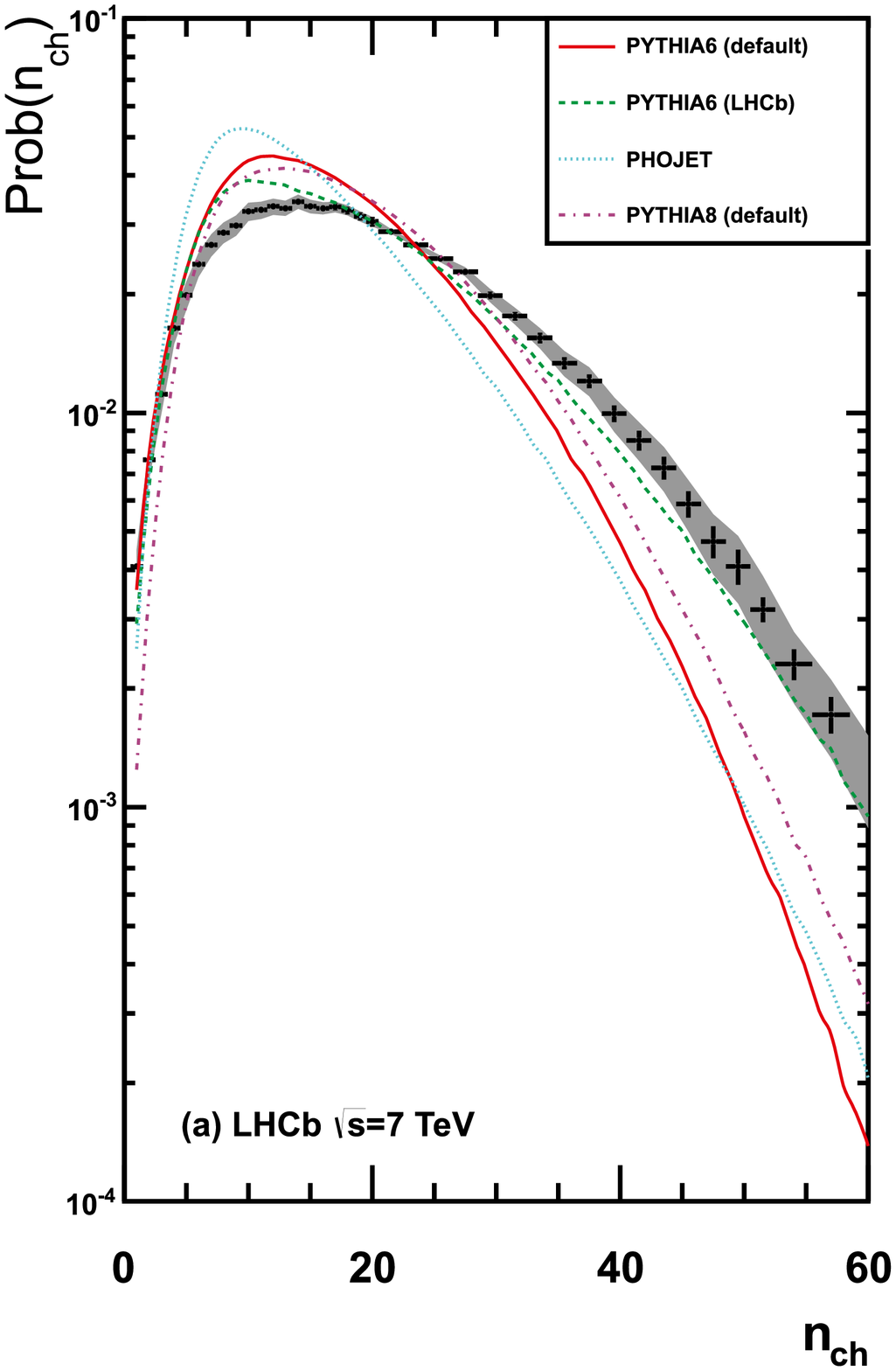}
\caption{Measured charged particle multiplicity in the forward range $2<\eta<4.5$ for the inclusive (left) and hard event sub-sample (right) compared to different generators.}\label{CPM_1}
\vspace{-10pt}
\hspace{-2pt}
\end{wrapfigure}
Further it allows to measure particle production in a small backward $\eta$ range ($-2.5<\eta<-2.0$) in addition to the regular forward coverage ($2<\eta<4.5$).
The measured multiplicity distribution (Fig. \ref{CPM_1}) for events with at least one track in the forward acceptance shows that all generators, namely PYTHIA 6, PYTHIA 8 and PHOJET underestimate the amount of promptly produced charged particles. However, testing PYTHIA 6 tunes for which diffractive processes were switched off at generator level seem to give an accidentally better agreement with the measured data. Studying a sub-sample of hard QCD events by requiring at least one high $p_{T}$-track with $p_{T}>1\;$GeV/c results in an increase of charged particles. At least for this hard event sample, some PYTHIA 6 tunes provide a reasonable description of the data.

\section{Prompt Hadron Ratios}
The LHCb collaboration measured prompt hadron production ratios~\cite{LHCb_PHR} as a function of pseudorapidity in three different $p_{T}$-bins for \textit{pp} collisions at centre-of-mass energies of $\sqrt{s}=0.9$ and $7\;$TeV. The measured anti-particle/particle ratios $K^{-}/K^{+}$, $\pi^{-}/\pi^{+}$ and $\overline p/p$ as well as the different-particle ratios $(p+\overline p)/(\pi^{+}+\pi^{-})$, $(K^{+}+K^{-})/(\pi^{+}+\pi^{-})$ and $(p+\overline p)/(K^{+}+K^{-})$ are probes for hadronisation models implemented in Monte Carlo event generators. Further, some of these ratios can be used to test models of baryon to meson and strangeness suppression. A crucial ingredient in measuring these ratios is a good particle identification which is provided by the two RICH detectors. The PID efficiencies were directly determined from data using decays of resonances like $\Lambda \rightarrow p\pi^{-}$, $\phi \rightarrow K^{+}K^{-}$ and $K_{S}^{0} \rightarrow \pi{+}\pi{-}$. The dominant systematic uncertainty remains the PID efficiency because of the limited size of the calibration sample.
Comparing the measured hadron ratios to different PYTHIA 6 tunes shows that no tune is able to describe the entire set of measurements. Only each type of hadron ratio can be described by at least one single tune. 
Of special interest is the $\overline p/p$ ratio, which is sensitive to baryon number transport. At $\sqrt{s}=0.9\;$TeV the $\overline p/p$ ratio has a significant $\eta$ dependence, which is qualitatively described by all PYTHIA 6 tunes. 
But only the Perugia NOCR tune, which favours an extreme model of baryon transport, is able to give a quantitatively good prediction while other generator tunes underestimate baryon transport. 
\begin{wrapfigure}{l}{0.5\textwidth}
\vspace{-5pt}
\includegraphics[width=0.5\textwidth]{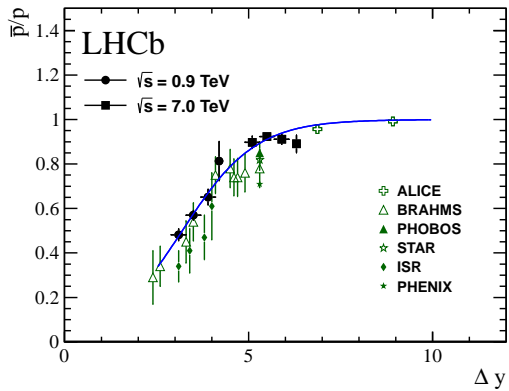}
\caption{Results for $\overline{p}/p$ ratio as function of rapidity loss. Fit to ALICE and LHCb data is superimposed.}\label{PHR_1}
\vspace{-5pt}
\end{wrapfigure}
However, at $\sqrt{s}=7\;$TeV the Perugia NOCR model tends to now overestimate baryon transport. The same ratio can also be studied as function of rapidity loss $\Delta y=y_{beam}-y_{particle}$, defined as the difference of the rapidity of the beam and the considered particles. This representation allows to compare measurements of experiments at different centre-of-mass energies, as it is shown in Fig. \ref{PHR_1}. The LHCb measurement covers a wider range in rapidity loss and improves previous measurements with a better precision. Combining the LHCb data points and the complementary ALICE measurement~\cite{Alice} allows to perform a fit within in the Regge model~\cite{Regge}. In this model, baryon production at high energies is driven by Pomeron exchange and baryon transport by string junction exchange. In this picture, the gained fit parameters are related to contributions from these two mechanisms. The fit result of a low string junction contribution with low intercept point allows to draw conclusions about the associated standard Reggeon or the Odderon.\\
 
% ****************************************************************************
% BIBLIOGRAPHY AREA
% ****************************************************************************

\begin{footnotesize}
% IF YOU DO NOT USE BIBTEX, USE THE FOLLOWING SAMPLE SCHEME FOR THE REFERENCES
% ----------------------------------------------------------------------------

\end{footnotesize}

\begin{thebibliography}{99}
%------- replace following references ;-)

\bibitem{LHCb} LHCb collaboration, A. A. Alves Jr. {\it et~al.}, JINST {\bf 3} S08005 (2008).
\bibitem{LHCb_EF} LHCb collaboration, R. Aaij {\it et~al.}, Eur. Phys. J. {\bf C73} 2421 (2013).
\bibitem{LHCb_CPM} LHCb collaboration, R. Aaij {\it et~al.}, Eur. Phys. J. {\bf C72} 1947 (2012).
\bibitem{LHCb_PHR} LHCb collaboration, R. Aaij {\it et~al.}, Eur. Phys. J. {\bf C72} 2168 (2012).
\bibitem{Alice} The ALICE collaboration, Phys. Rev. Lett. {\bf 105} 072002 (2010).
\bibitem{Regge} D. Karzeev, Phys. Lett. B {\bf 378} 238 (1996).

\end{thebibliography}
\end{document}